# System–Technology Co-Optimization of Bitline Routing and Bonding Pathways in Monolithic 3D DRAM Architectures

Kiseok Lee, Sungwon Cho, Seongkwang Lim, Suman Datta and Shimeng Yu
School of Electrical and Computer Engineering, Georgia Institute of Technology, Atlanta, Georgia, USA
Email: shimeng.yu@ece.gatech.edu

***Abstract*—3D DRAM has emerged as a promising approach for continued density scaling, but its viability is limited by routing and hybrid bonding constraints to periphery, which may degrade sensing margin, latency, and array efficiency. With device characteristics and array parasitics extracted from TCAD, SPICE simulations are performed with peri logic in a CMOS-Bonded-Array (CBA). The analysis shows that the bitline strap architecture with amorphous oxide semiconductor (AOS) selectors is essential to manage routing congestion and parasitics. The optimized design achieves a bit density of 2.6 Gb/mm$^2$ (137 layers with Si access transistors or 87 layers with AOS), representing ~6× density scaling over D1b 2D DRAM. The design further demonstrates a nominal row cycle time (tRC) of 10.5 ns, compared to 21.3 ns in D1b, and a 60% reduction in read/write energy.**

## I. Introduction

3D DRAM is key to sustaining density scaling beyond $4F^2$ [1]. Recent works prototyped this technology and analysis so far focused on the bit cell and mat (sub-array) level with discussions on access transistor characteristics and tradeoffs between vertical BL and vertical WL [1-6]. Even VBL is preferred, the vertical extension of BLs and the stacking of WLs create severe routing congestion, increased parasitics, and vulnerability to disturb mechanisms such as floating-body effects (FBE) and row-hammering (RH) [7, 8]. To our best knowledge, the analysis beyond the mat-level is lacking for 3D DRAM, while the employment of hybrid bonding introduces new challenges with global routing parasitics and peripheral circuit placement [7]. This study focuses on the architectural co-optimization that was missing in prior works (Table I), specifically targeting BL routing topologies and CMOS-Bonded-Array (CBA) integration.

## II. Methods and Results

We performed TCAD parasitic extraction and access transistor modeling of Si and amorphous oxide semiconductor (AOS) channel cell architectures, with the AOS model calibrated against the low-leakage W-doped $In_2O_3$ (IWO) double gate transistor [9]. The storage node capacitance (Cs) was unified at 4 fF, matching the D1b [10] estimate. The optimal cell configuration was selected; dimensions, extracted characteristics and parasitics are summarized in Fig. 1. The proposed cell architecture features line-type cell isolation (iso) and Si deposition-based mold strategy. Two distinct iso methods are evaluated: Contact-type iso [2, 4] utilizes contact-etch for bilateral WL patterning followed by isotropic Si etching to complete cell iso. However, the lateral recess typically increases the Y-pitch or constricts the channel width (40nm). Following a refined design, line-type iso [7] initiates with a line-etching process and proceeds with full gate wrapping around the Si channel. This configuration minimizes Y-pitch (100nm) while achieving a wider channel width (70 nm). Furthermore, Si deposition-based mold (with channel last and inner contact) for AOS is adopted to effectively reduce the iso-etch pitch [7]. This is enabled by the ease of Si etching compared to $SiO_2$ and $Si_3N_4$, facilitating more aggressive scaling. BL routing is the dominant factor limiting sense margin and array efficiency.

TABLE I
Comprehensive benchmarking of 1T1C 3D DRAM technologies: A multi-level comparison of "This Work" with state-of-the-art implementations from cell and array to architecture and system performance metrics.

| | 1T1C 3D DRAM | Samsung VLSI 23 [1] | SK Hynix VLSI 24 [2] | CAS,SAMT IEDM 24 [3] | imec VLSI 25 [4] Imec IEDM 25 [5] | Kioxia IEDM 25 [6] | This Work (What is new?) |
|---|---|---|---|---|---|---|---|
| Cell | Cell structure | double gate | double gate contact type isolation | P-CAA Isolated channel | GAA contact type isolation | GAA separated WL / gate | GAA line type isolation |
| | Channel material (mold) | epitaxial Si (Si-SiGe) | epitaxial Si (Si-SiGe) | AOS (TiN-$SiO_2$) | epitaxial Si (Si-SiGe) | AOS ($SiO_2$-$Si_3N_4$) | epitaxial Si (Si-SiGe) & AOS (Si deposition) |
| | Access Tr. analysis | experiment + TCAD | experiment | experiment | experiment + TCAD | experiment | TCAD |
| Array | Array direction | VBL vs. HBL | VBL | HBL | VBL | VBL | VBL |
| | Array characteristics | 6 tier demonstration (no quantitative result) | 5 tier demonstration (no quantitative result) | 4 tier demonstration (no quantitative result) | 3 tier demonstration (no quantitative result) | 8 tier demonstration (no quantitative result) | TCAD mini array→ full array estimation |
| | Vulnerability | simple mention : FBE | experiment : FBE | experiment : BTI, endurance, retention | experiment : FBE | experiment : BTI, endurance, retention | TCAD : FBE + RH |
| Architecture & System | Array organization WL/BL routing to CMOS CMOS formation | simple mention : HCB CBA | experiment : HCB CBA | simple mention : BL selector, strap | none | simple mention : BL/WL selector,strap | HCB CBA : BL/WL selector, strap layout & quantitative analysis (TCAD+SPICE) |
| | Bit density | none | simple mention (a.u.) | D0a ~72L | D0a 100L, D0b 150L, D0c 200L | simple mention (a.u.) | D0f 2.6Gb/mm$^2$: 137L (Si), 87L (AOS) |
| | Sense margin | none | initial △VBL | initial △VBL | none | initial △VBL (a.u.) | full SWD+BLSA circuit (SPICE, compact model) |
| | Latency (timing) | none | none | simple mention (a.u.) | none | none | |
| | Energy efficiency | none | none | none | none | simple mention (a.u.) | analytical estimation |

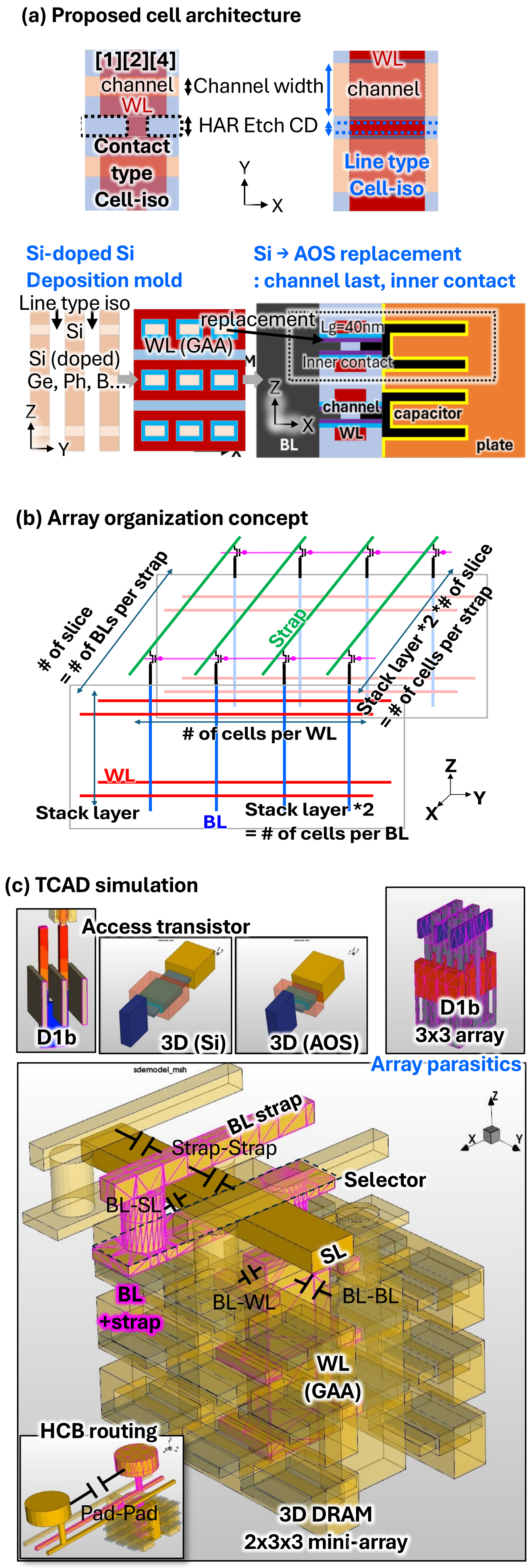

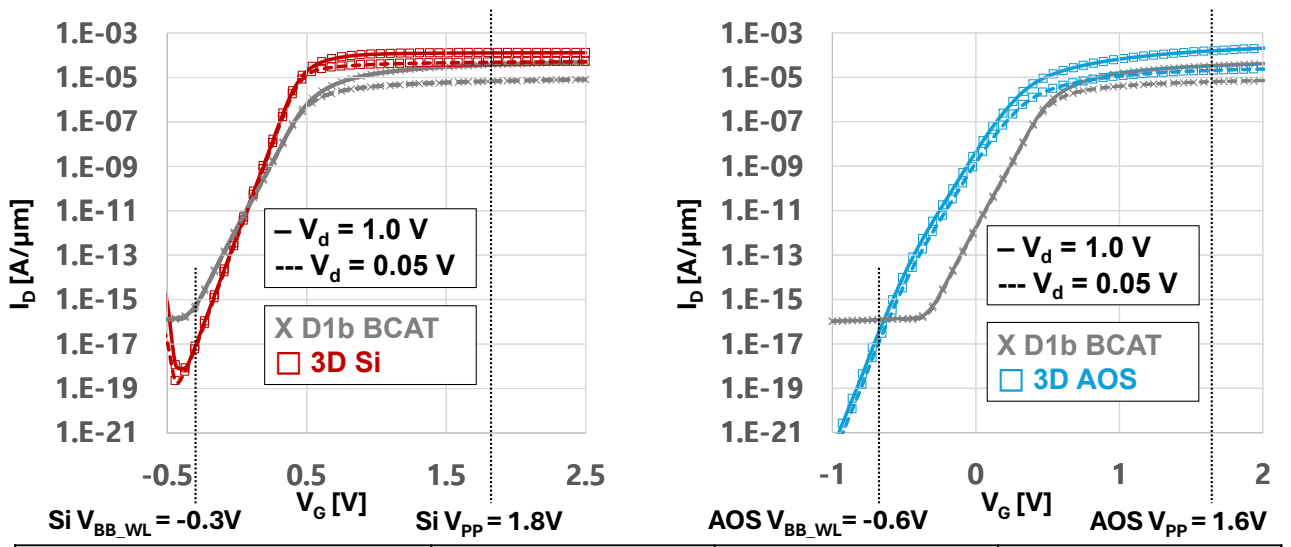

| VBL, CBA, selector | | 2D (D1b [11]) | 3D (Si) | 3D (AOS) |
|---|---|---|---|---|
| Access transistor | | BCAT | GAA line type cell isolation | |
| Channel | | c-Si | Epitaxial Si | IWO |
| Dimension | X / Lg [nm] | 32.6 / 120 | 349 / 100 | 238 / 40 |
| | Y / W [nm] | 37.6 / 11.7 | 100 / 70 | 100 / 70 |
| | Z [nm] | - | 70 | 80 |
| # of Cells per BL / WL | | 1280 / 1024 | #of layer*2 / 1024 | #of layer*2 / 1024 |
| Cs [fF] | | 4 | | |
| $C_{BL}$ per 1layer [fF] | | 25.0 | 0.0815 | 0.128 |
| $R_{BL}$ per 1layer [kΩ] | | 49.6 | 0.292 (N+ p-Si) | 0.0167 (TiN/W) |
| $C_{WL}$ [fF] / $R_{WL}$ [kΩ] | | 30.0 / 81.2 | 96.3 / 8.1 | 94.4 / 19.9 |
| Parasitic $C_{WL}$ [fF] | | 16.2 | 42.0 | 33.2 |
| Ion [μA] / Ioff [fA] | | 2.44 / 0.2 | 9.03 / 0.02 | 10.4 / 0.02 |
| Vth / $V_{PP}$ [V] | | 0.43 / 2.5 | 0.30 / 1.8 | 0.20 / 1.6 |
| $V_{BB_WL}$ / $V_{BB}$ [V] | | -0.3 / -0.6 | -0.3 / - | -0.6 / - |

Fig. 1. Geometric and electrical characteristics of baseline D1b [10] and proposed 1T1C 3D DRAM (a) proposed cell architecture and integration strategy, (b) array organization concept, (c) multi-scale TCAD simulation (access transistor, array, routing) and results.

We evaluated four distinct routing schemes: (a) Direct BLSA connection, (b) BL strapping, (c) Core MUX, and (d) BL Selector + Strap (Fig. 2). While schemes (a) and (c) offer the lowest delay, they demand a prohibitively tight hybrid Cu bonding (HCB) pitch of 0.26 μm (Si) and 0.22 μm (AOS). Simple strapping (b) relaxes pitch requirements but substantially increases BL parasitic capacitance (CBL), thereby degrading the sense margin. The proposed solution is the "BL Selector + Strap".

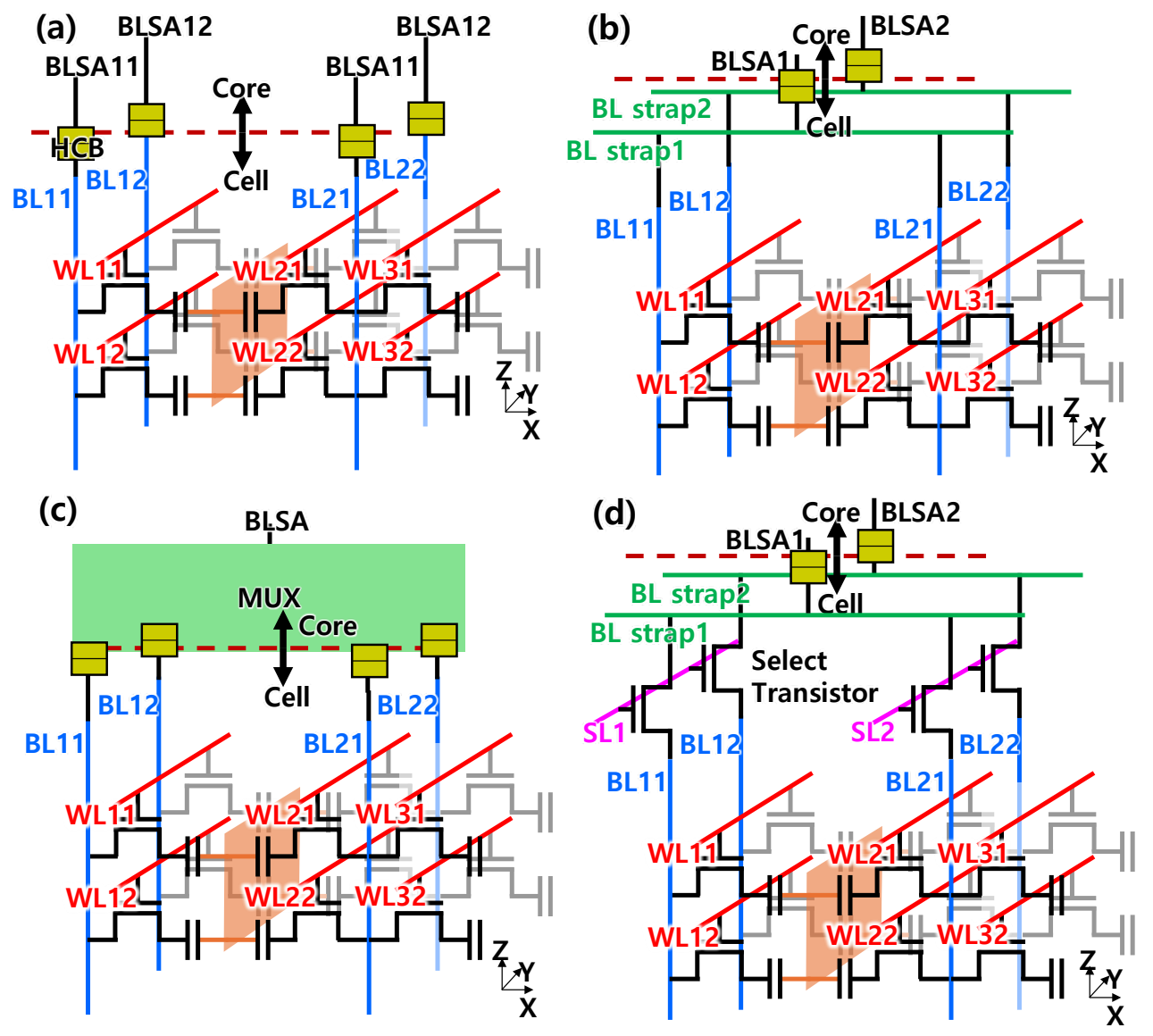

Fig. 2. Schematic comparison of four BL routing architectures. (a) Direct BLSA connection, (b) BL strapping, (c) Core MUX, and (d) BL Selector + Strap.

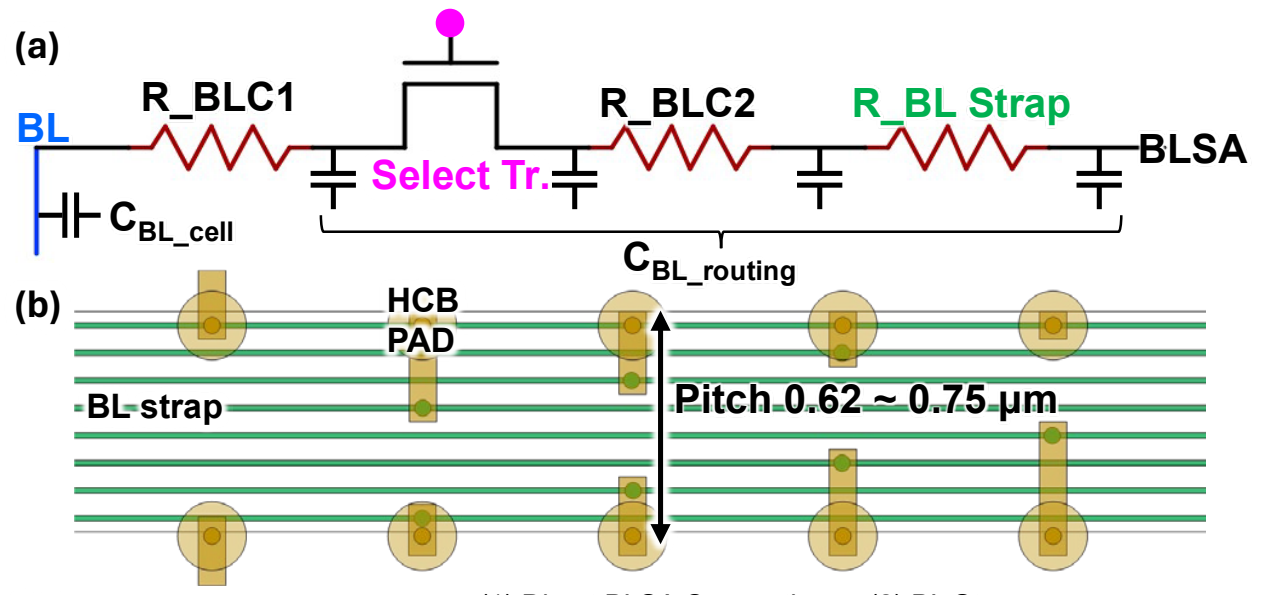

(c)

(1) Direct BLSA Connection (2) BL Strap
(3) BL Multiplexing (Core) (4) BL Selector + Strap (Cell)

| Assumption: D1b 1024 Cells /BL 3D 1024 Cells /Strap 1024 Cells /WL Cs 4fF Risk | D1b [10] | 3D (Si) | | | | 3D (AOS) | | | |
|---|---|---|---|---|---|---|---|---|---|
| | | (1) | (2) | (3) | (4) | (1) | (2) | (3) | (4) |
| $C_{BL}$ [fF] total | 20 | 5.3 | 25 | 5.3 | 6.6 | 5.5 | 34 | 5.5 | 6.6 |
| └$C_{BL}$ [fF] routing | | <0.1 | 1.6 | <0.1 | 1.3 | <0.1 | 1.4 | <0.1 | 1.1 |
| Initial $\Delta V_{BL}$ [mV] | 70 | 215 | 69 | 215 | 189 | 211 | 53 | 211 | 189 |
| $R_{BL}$ [kΩ] total | 40 | 19 | 23 | 23 | 23 | 0.7 | 3.4 | 3.4 | 3.4 |
| └$R_{BL}$ [kΩ] routing | | <0.1 | 3.7 | <0.1 | 3.7 | <0.1 | 2.7 | <0.1 | 2.7 |
| HCB pitch [μm] | | 0.26 | 0.75 | 0.26 | 0.75 | 0.22 | 0.62 | 0.22 | 0.62 |
| BLSA/SWD area [μm²] | 0.44/0.17 | 0.14 | 1.12 | 1.12 | 1.12 | 0.10 | 0.76 | 0.76 | 0.76 |
| BL disturb count [ratio] | 1 | 1 | 1 | 1/8 | 1/8 | 1 | 1 | 1/8 | 1/8 |
| Sense margin [mV] | 54 | Fail | Fail | 148 | 130 | 210 | 53 | 210 | 189 |

Fig. 3. (a) equivalent circuit, (b) layout of BL strap and HCB routing and (c) comprehensive comparison of BL routing schemes.

As shown in Fig.3, this configuration isolates unselected BLs, reducing effective CBL to 6.6 fF (w/ bonding parasitics) (vs. 20 fF for D1b) and improving the sense margin to 130 mV (Si) and 189 mV (AOS) (vs. 54 mV for D1b). Furthermore, the select transistor enables the inactive BL to float at a refresh potential, effectively mitigating the FBE and off-state leakage by decoupling the cell from the global BL. This strategy also accommodates a relaxed HCB pitch of 0.75 μm (Si) and 0.62 μm (AOS), which is well within the manufacturable window for W2W HCB, while allowing for a larger BLSA area of 1.12 μm² (Si), 0.76 μm² (AOS) compared to D1b (0.44 μm²).

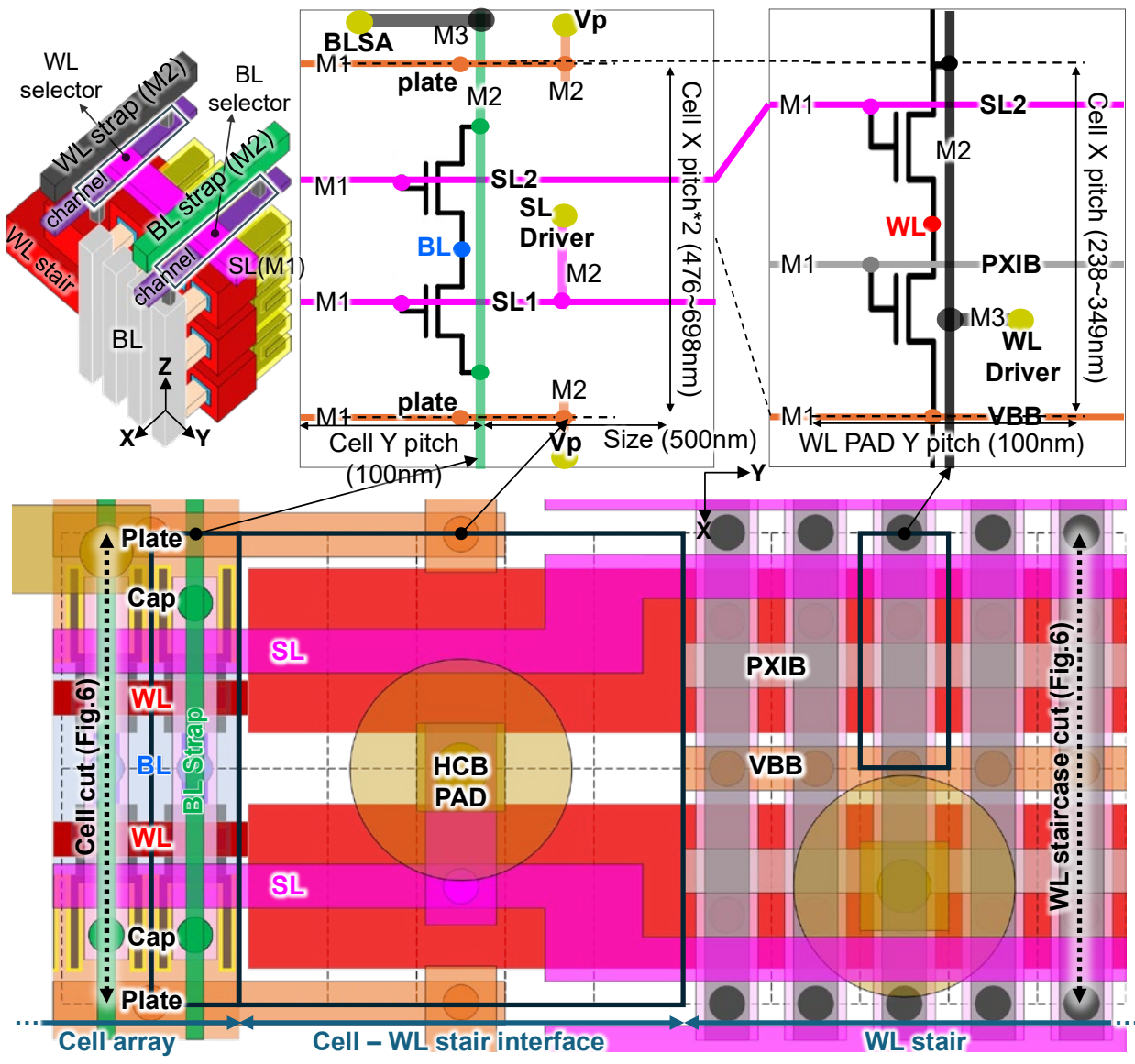

Fig. 4. Layout of selector, WL driver, SL driver and Vp routing.

The physical implementation of this architecture employs a CBA, where the periphery is vertically bonded over the cell array (Fig. 4-5). As detailed in the layout (Fig. 4), the design utilizes a metal stack (M1-M3) to route select lines (SLs) and WL to their drivers within the tight cell pitch, ensuring direct vertical connectivity. The layout (Fig.5) integrates 16 WLs and 8 BLs per strap, routing them to drivers and BL sense amplifiers (BLSA) located on the overlying CMOS wafer. This vertical integration effectively eliminates the long lateral routing bottlenecks inherent to 2D designs.

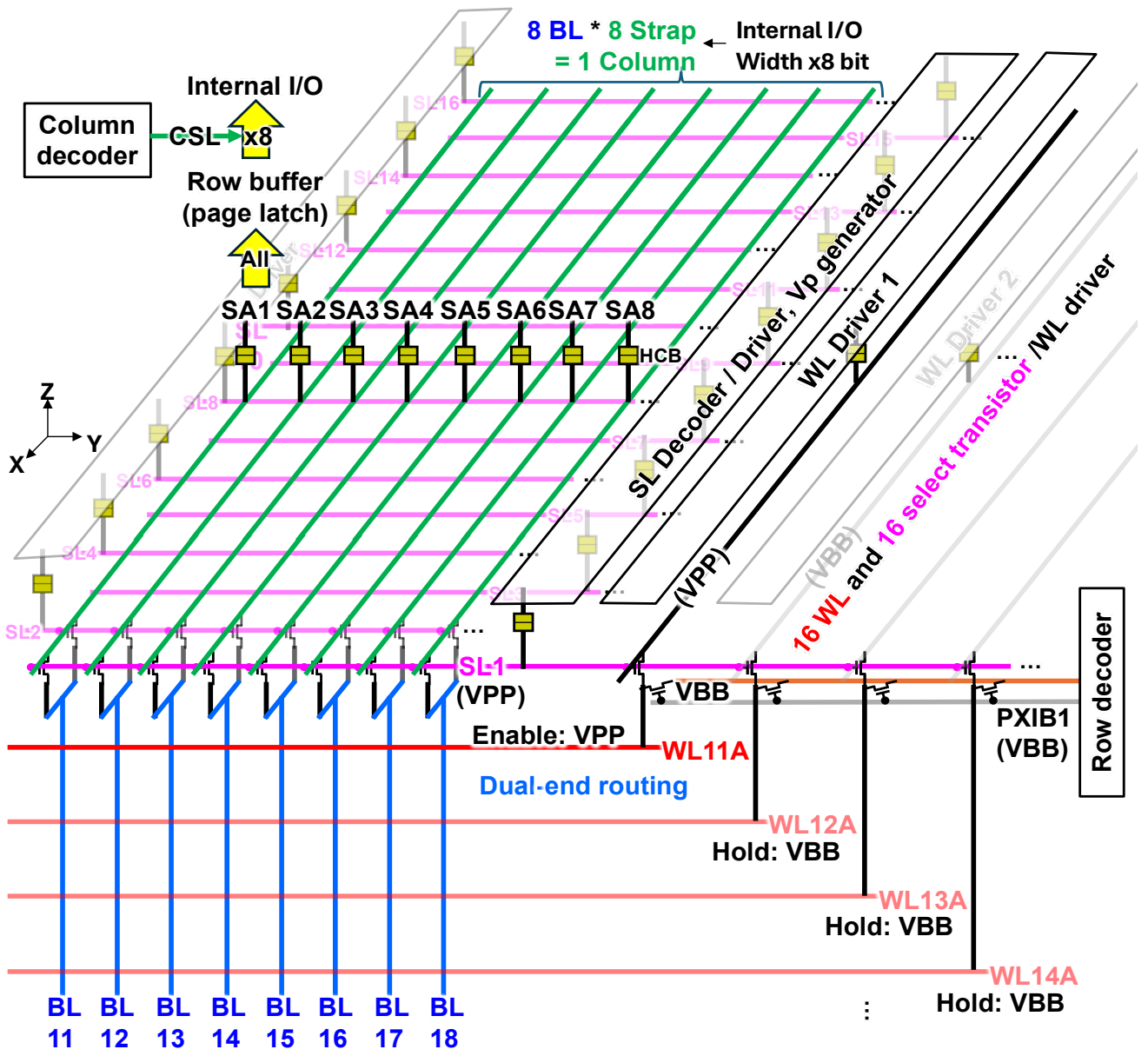

Fig. 5. WL and BL routing, SL/WL driver, BLSA and data path.

As shown in Fig. 6, select transistors are integrated on top of the cell to alleviate routing congestion. To ensure high drivability within tight spatial constraints, Indium-Gallium-Oxide (IGO [11]) was adopted for its performance and BEOL compatibility. The IGO selector can achieve Ion > 50μA @2V (W/L = 70 nm / 50 nm) and a near-ideal 60 mV/dec SS. Such a high-performance device is critical for 16 WL and 8 BL multiplexing schemes, ensuring efficient signal delivery while maintaining a compact footprint.

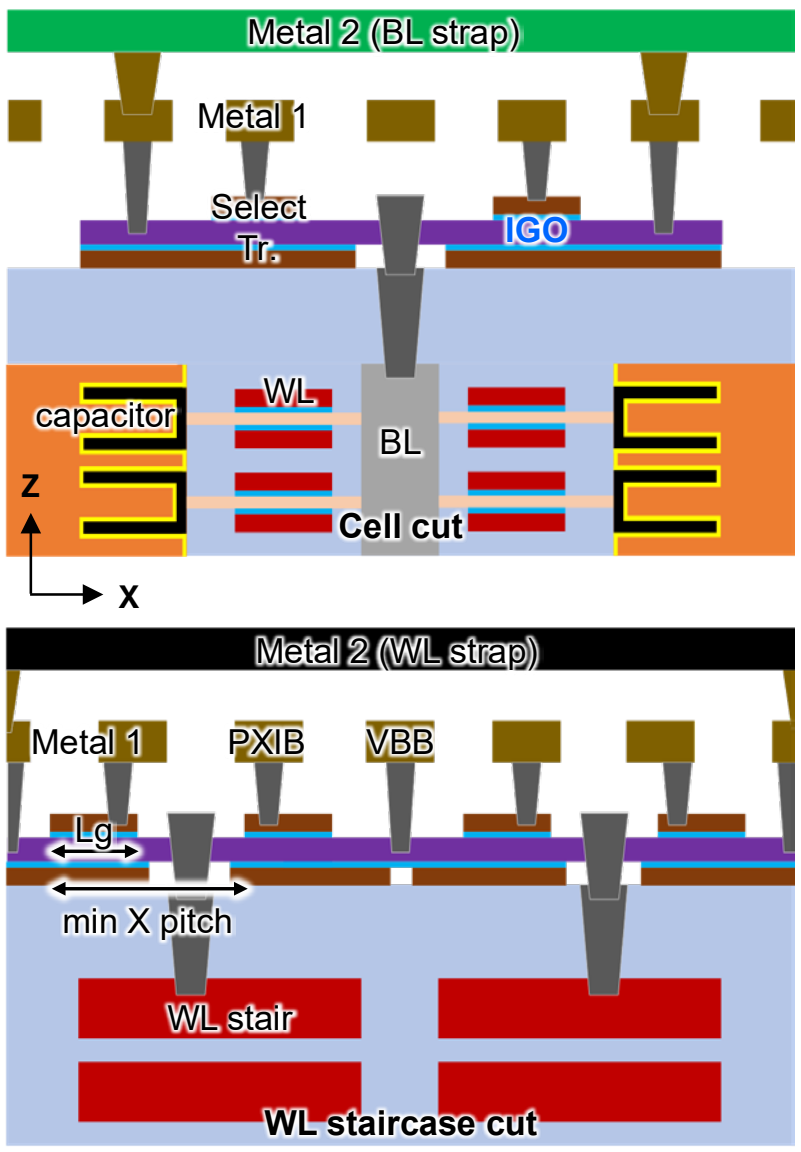

| 8 BL /Strap 16 WL /Strap | BL/WL select Tr. |
|---|---|
| Channel | IGO BEOL compatible |
| X pitch [nm] | min 119 @AOS case |
| →Gate length [nm] | 50 |
| Structure | DG |
| Y pitch [nm] | 100 |
| →Gate width [nm] | 70 |
| Ion [μA] | >50 @2V |
| SS [mV/dec] | approximately 60 |
| On-off ratio | $>10^{11}$ |

Fig. 6. Cross-section view and performance benchmark of the proposed IGO-based select transistor.

Fig. 7 outlines SPICE circuit topology and operating conditions; notably, the 3D architecture operates at reduced WL overdrive voltages (VPP: 1.6–1.8V), optimizing power consumption. Fig. 8 presents the transient simulation results for the full access path. Enabled by minimized parasitic RC delay, the proposed 3D DRAM achieves a nominal row cycle time (tRC) of <10.9 ns (Si), <10.5 ns (AOS), significantly faster than the 21.3 ns of the D1b [10] (Fig.9(c)). Disturbance-induced charge loss was analyzed via mixed-mode TCAD, assuming 10k RH toggles and $1.5 \times 10^6$ tRC cycles per 64ms. Transient spikes indicate BL disturb (FBE) and WL coupling (RH). Scaling projections in Fig.9(a) show that achieving a bit density of 2.6 Gb/mm$^2$ requires 137 layers (height = 9.6 μm) for Si and 87 layers (height = 6.9 μm) for AOS. As bit density increases, Fig.9(b) illustrates the expected reduction in sense margin due to increased parasitics with FBE and RH-induced charge loss; however, the Si case maintains functional margins (70 mV) at the 2.6 Gb/mm$^2$ target. Energy efficiency is similarly improved; write energy is reduced to 6.26/5.38 fJ (Si/AOS), read energy to 1.57/1.35 fJ, consistently surpassing the D1b across all performance metrics (Fig.9(c)).

| Voltage [V] | D1b | 3D(Si) | 3D(AOS) |
|---|---|---|---|
| NWEIB | 2.5~0 | 1.8~0 | 1.6~0 |
| PXID | 0~2.5 | 0~1.8 | 0~1.6 |
| PXIB | 1.1~0 | 1.1~0 | 1.1~0 |
| VPP | -0.3~2.5 | -0.3~1.8 | -0.6~1.6 |
| VSL | - | -0.6~2.5 | |
| VDD/VSS | 1.1 / 0 | | |
| LA | 1.1~0 | | |
| LAB | 0~1.1 | | |
| EQ | 0~1.4 | | |

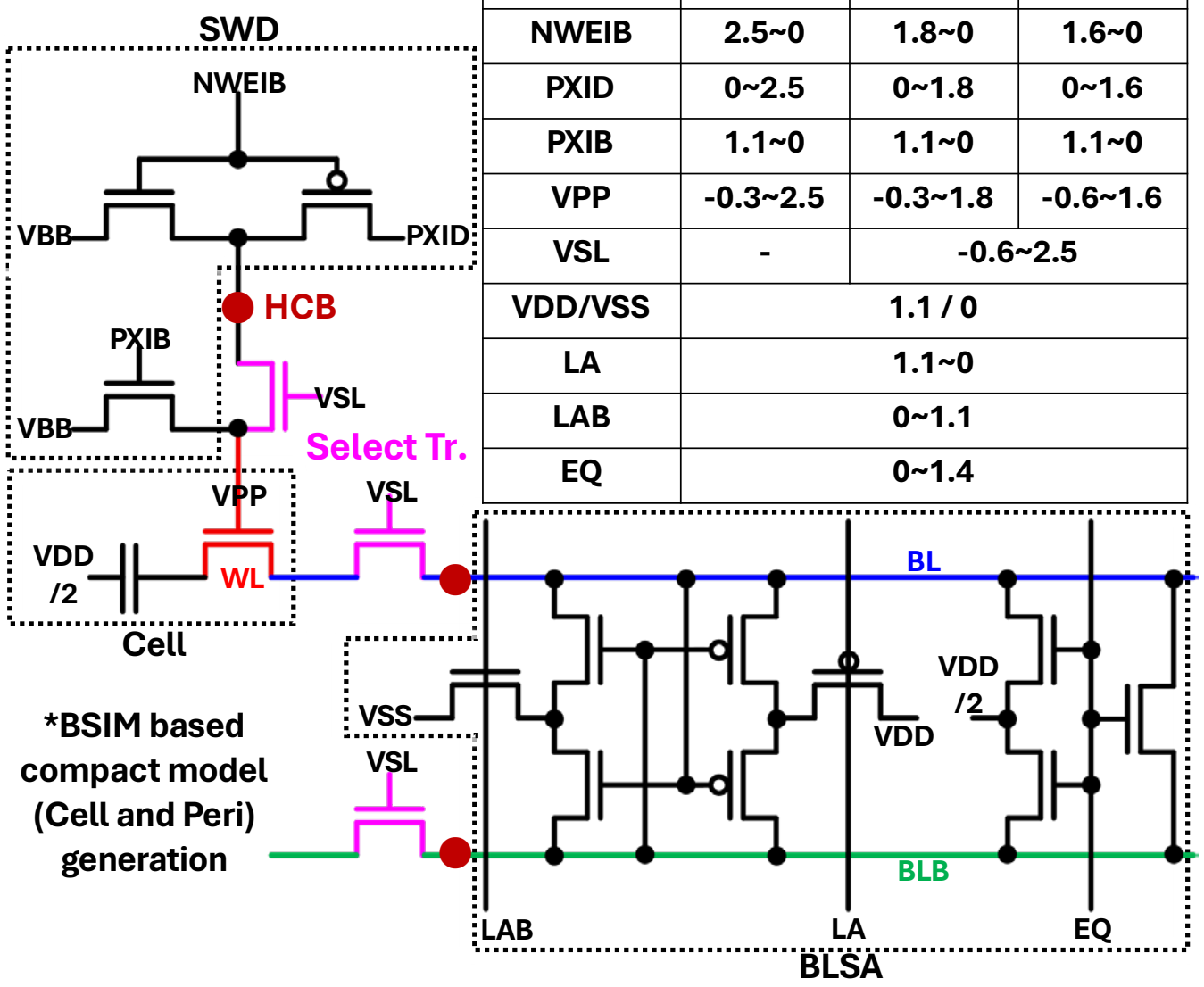

Fig. 7. Circuit diagrams utilized for SPICE simulations. The inset table summarizes the operating voltage conditions.

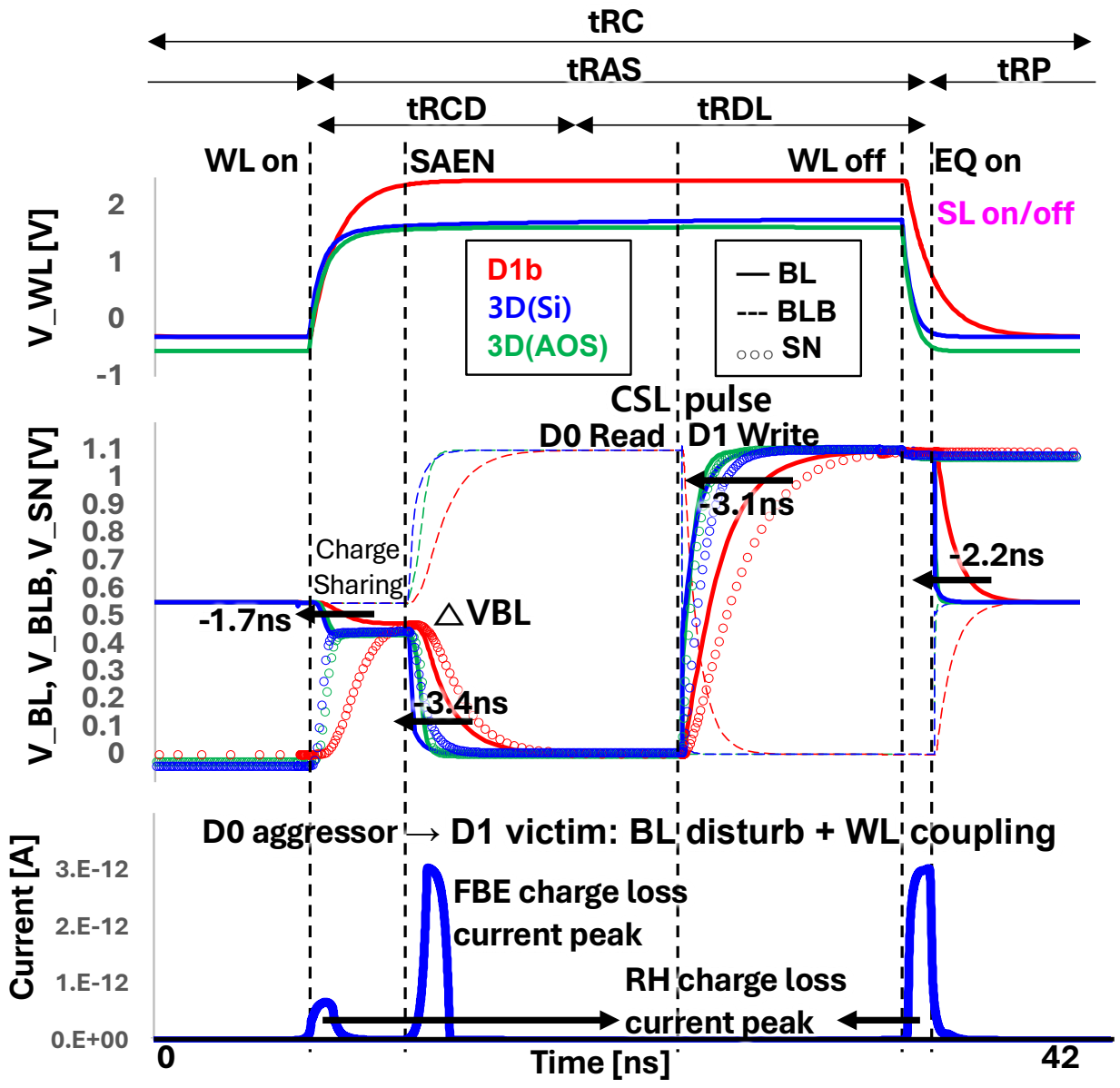

Fig. 8. SPICE/TCAD transient simulation waveforms of the full row cycle (tRC, fixed 42ns, 3D DRAM 2.6 Gb/mm$^2$) operation.

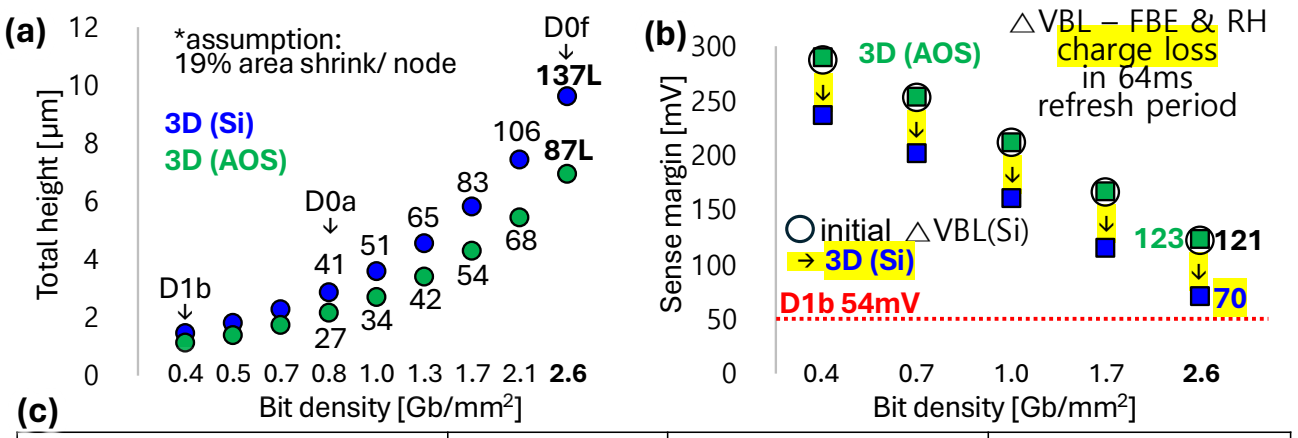

| D1b 0.436 [10], 3D 2.6 Gb/mm² | | D1b | 3D (Si) 137L | 3D (AOS) 87L |
|---|---|---|---|---|
| WL routing | | Single-end | Dual-end, selector, 16WL /strap | |
| BL routing | | - | Selector, 8 BL /Strap | |
| tRC, nominal [ns] | | 21.3 | < 10.9 | < 10.5 |
| └charge sharing 95% [ns] | | 2.9 | 1.2 | 1.3 |
| └SA amplifying 95% [ns] | | 4.5 | 1.1 | 1.1 |
| └write 95% [ns] | | 4.9 | 1.8 | 1.3 |
| └pre charge 95% [ns] | | 4.0 | 1.8 | 1.8 |
| └CSL pulse [ns] | | 5.0 | < 5.0 | < 5.0 |
| Energy /cell (~BLSA) | Write [fJ] | 15.50 | 6.26 | 5.38 |
| | Read [fJ] | 3.88 | 1.57 | 1.35 |

Fig. 9. Scaling projections: (a) total stack height (b) sense margin vs. bit density and (c) comprehensive comparison of architectural specifications and performance metrics (fixed 2.6 Gb/mm$^2$).

## III. Conclusion

Our analysis confirms that the selector-strap topology effectively minimizes parasitics and vulnerability, securing robust sensing margins and manufacturable bonding pitches. 3D DRAM with CBA technology improves latency and energy efficiency over 2D baselines.

## IV. Acknowledgments

This work is supported by PRISM, one of the SRC/DARPA JUMP 2.0 centers.